\documentclass[twocolumn,         
               showpacs,longbibliography,            
               showkeys,preprintnumbers,     
               aps,                 
               prd,          	    
               letterpaper,             
               superscriptaddress,      
               nofootinbib,         
               tightenlines,        
               floats,floatfix      
               ]{revtex4-1}
\usepackage{physics}
\usepackage{epsf}
\usepackage{graphicx,color}
\usepackage{subfigure}
\usepackage{latexsym}
\usepackage{amsmath,amssymb}        
\usepackage[colorlinks=true,linkcolor=blue,citecolor=blue]{hyperref}
\usepackage{mathrsfs}
\usepackage{comment}
\usepackage{soul}
\usepackage{cancel}
\usepackage[normalem]{ulem}
\definecolor{purple}{rgb}{0.58,0.0,0.83}
\definecolor{orange}{rgb}{1,0.5,0}
\DeclareSymbolFontAlphabet{\mathrsfs}{rsfs}
\DeclareMathAlphabet{\mathcal}{OMS}{cmsy}{m}{n}

\begin{document}


\title{Galactic Rotation Curves of LSB Galaxies using core-halo FDM configurations}


\author{Iv\'an  \'Alvarez-Rios}
\email{ivan.alvarez@umich.mx}
\affiliation{Instituto de F\'{\i}sica y Matem\'{a}ticas, Universidad
              Michoacana de San Nicol\'as de Hidalgo. Edificio C-3, Cd.
              Universitaria, 58040 Morelia, Michoac\'{a}n,
              M\'{e}xico.}

\author{Tula Bernal}
\email{tbernalm@chapingo.mx}
\affiliation{\'Area de F\'isica, Depto.~de Preparatoria Agr\'icola, Universidad Aut\'onoma Chapingo, Km 38.5 Carretera M\'exico-Texcoco, Texcoco 56230, Edo. M\'ex., M\'exico}             

\author{Pierre-Henri Chavanis}
\email{chavanis@irsamc.ups-tlse.fr}
\affiliation{Laboratoire de Physique Th\'eorique, Universit\'e Paul Sabatier,
 118 route de Narbonne 31062 Toulouse, France}             

\author{Francisco S. Guzm\'an}
\email{francisco.s.guzman@umich.mx}
\affiliation{Instituto de F\'{\i}sica y Matem\'{a}ticas, Universidad
              Michoacana de San Nicol\'as de Hidalgo. Edificio C-3, Cd.
              Universitaria, 58040 Morelia, Michoac\'{a}n,
              M\'{e}xico.}  


\date{\today}


\begin{abstract}
In this work, we construct galactic halos in order to fit the rotation curves (RCs) of a sample of low surface brightness (LSB) galaxies. These halos are made of Fuzzy Dark Matter (FDM) with a multimode expansion of non-spherical modes that in average contribute to the appropriate density profile consisting of a core and an envelope needed to fit the rotation curves. The coefficients of the expansion are calculated using a genetic algorithm, that minimizes the difference between the spatial average density of the multimode order parameter describing the FDM and the target dark matter density that fits the RCs. The FDM halos are constructed assuming a solitonic core at the center and two types of envelopes, Navarro-Frenk-White and Pseudo-Isothermal density profiles. The resulting FDM configurations are then evolved in order to show how the average density changes in time due to the secular dynamical evolution, along with a condensation process that lead to the growth of the solitonic core.
\end{abstract}




\maketitle

\section{Introduction}
\label{sec:intro}

Fuzzy Dark Matter (FDM) is a dark matter candidate, consisting of an ultralight spin zero boson that has received recent attention because it apparently solves some of the traditional problems of Cold Dark Matter (CDM), namely the cusp-core and the too-big-to-fail problems as explained in recent reviews \cite{Suarez:2013,Hui:2021tkt,ElisaFerreira,Chavanis2015,Niemeyer_2020}. The reason is that the formation of very small scale structures is prevented by the uncertainty principle for such an ultralight particle and the mass power spectrum is cut-off at small scales. In addition, the tiny mass  of the boson implies smooth galactic cores as opposed to the cuspy shape obtained from predictions of CDM.

At cosmic scales the model has been deeply studied in structure formation simulations (SFS) (see e.g. \cite{Schive:2014dra,Mocz:2017wlg,Veltmaat_2018,MoczPRL2019,May_2021,Gotinga2022}), that are promising and already involve the dynamics of baryonic matter. At local scales the works concentrate on the formation of core-tail halos like those obtained in SFS, for example through the merger of multiple cores (see e. g. \cite{Schive:2014hza,Mocz:2017wlg,Schwabe:2016}) that end-up with the core surrounded by a typical granular structure that in average shows a Navarro-Frenk-White (NFW) density profile \cite{NFW}. Construction of target density profiles is also a subject of current interest, because the wave function describing the FDM at local scales suggests a clear multimode dependency. This approach has been developed for SFS \cite{PhysRevD.97.103523} as well at local scales with the construction of on demand multimode density profiles whose stability is studied with simulations \cite{YavetzLiHui2022,Luna2023b}.

Now, the boson mass $m_B$ in the FDM, in order to address the small scale problems (core density profile and suppression of the small-scale structure) and to behave like CDM on large scales, must be of the order of $m_B \sim 10^{-23}-10^{-21}$ eV. From the high-redshift luminosity function of galaxies we have the constraint for the boson mass $m_B > 1.2 \times 10^{-22}$ eV \citep{Schive:2015}, while  \citet{Irsic:2017,Armengaud:2017} derive a stringent constraint, indicating $m_B \gtrsim 2 \times 10^{-21}$ eV. On the lower limit of the boson mass, the most used value is $m_B \sim 10^{-22}$ eV in order to solve the small-scale problems of CDM. In the cosmological context, the analysis of Cosmic Microwave Background (CMB) and galaxy clustering data in e.g. \citet{Hlozek:2014}, establishes a constraint for the boson mass in the FDM model of $m_B > 10^{-24}$ eV. Considering the galaxy UV-luminosity function and reionization constraints, \citet{Bozek:2014} determined a minimum mass requirement of $m_B > 10^{-23}$ eV. Also, from Lyman-$\alpha$ observations, the constraint is $m_B > 10^{-23}$ eV \citep{Sarkar:2016}. This value is in tension with the results by \citet{Rindler-Shapiro:2012}, setting the minimum value for $m_B > 10^{-25}$ eV. This indicates there is no consensus on the accurate mass of the ultra-light boson and that further exploration is still necessary. Meanwhile we explore the viability of the model at local scales. Notice that self-interaction is another parameter that influences the construction and phenomenology of structures within the bosonic dark matter model, and that it could substantially change the constraints on the boson mass \cite[see e.g.][]{Robles:2012kt,RindlerDaller:2013,Suarez:2016eez,Suarez:2017mav,Desjacques:2017fmf,Cembranos:2018ulm,Urena-Lopez:2019xri,Chavanis:2020rdo,Delgado:2022vnt}.

In this work, likewise in \cite{Luna2023b}, we focus on the construction of multimode FDM configurations, in particular with solitonic core and an envelope with NFW and Pseudoisothermal (PISO) density profiles that fit rotation curves of low surface brightness (LSB) galaxies, and study their evolution in order to study their behavior and stability properties.

The article is organized as follows. In Section \ref{sec:core-halo} we describe the method we use to construct multimode halos, in Section \ref{sec:stability} we study the evolution of these configurations, and finally in Section \ref{sec:conclusions} we draw some conclusions.

\section{Construction of galactic core-halo profiles}
\label{sec:core-halo}

\subsection{Basic assumptions}
\label{sec:ba}

The dynamics of FDM is modeled with the Schr\"odinger-Poisson (SP) system:

\begin{equation}
i\hbar\frac{\partial\Psi}{\partial t} = -\frac{\hbar^2}{2m_B}\nabla^2\Psi + m_B V \Psi,
\label{eq:Gross-Pitaevskii}
\end{equation}

\begin{equation}
\nabla^2 V = 4\pi G \left(\rho - \bar{\rho}\right),
\label{eq:Poisson}
\end{equation}

\noindent where $\Psi$ is an order parameter related to the matter density through $\rho := m_B |\Psi|^2$, with $m_B$  the boson particle mass, $\hbar$ the reduced Planck constant, $G$ the gravitational constant, and $\bar{\rho} = \frac{1}{|D|}\int_D \rho \,d^3x$ the spatially averaged density calculated within a spatial domain $D$ with volume $|D| := \int_D d^3x$, where the construction of configurations is implemented and where the evolution is carried out. The gravitational potential $V$ is sourced by the difference between the density and its spatial average.

We want to construct solutions of the Schr\"odinger-Poisson (SP) system that are consistent with some galactic rotation curves. In order 
to construct the wave function of the core-halo, we follow a similar strategy as that designed in \cite{PhysRevD.97.103523} and \cite{YavetzLiHui2022}. We assume there is a target density profile $\rho_{T}$, and the goal is to construct a corresponding wave function $\Psi_0$ that is consistent with this density profile and satisfies the SP system. For this, we consider the target density to be a spherically symmetric function, depending on the radial coordinate $r$ only. This makes possible to solve Poisson equation (\ref{eq:Poisson}) in spherical symmetry which can be written as the following first order system:

\begin{eqnarray}
&& \dfrac{d V_T}{dr}= G\dfrac{M_T}{r^2}, \label{eq:stationaryGPPa}\\
&& \dfrac{d M_T}{dr} = 4\pi r^2 \rho_T, \label{eq:stationaryGPP}
\end{eqnarray}

\noindent where $\rho_T$ is the target density and $M_T$ is the radial mass function. Once Poisson equation is solved, the resulting potential $V_T$ is a function of $r$. This potential is injected into the stationary version of the Gross-Pitaevskii equation (\ref{eq:Gross-Pitaevskii}). This equation is reminiscent of the problem of the hydrogen atom, with the notable difference being the replacement of the Coulomb potential by the potential $V_T$, which is written as a Sturm-Liouville problem:

\begin{equation}
 -\frac{\hbar^2}{2 m_B}\frac{1}{r^2}\frac{\partial}{\partial r}\left(r^2\frac{\partial\psi_{j}}{\partial r}\right) +\frac{\hbar^2}{2m_B}\frac{L^2}{r^2}\psi_j+ m_B V_T\psi_{j} = E_{j}\psi_{j}, \label{eq:stationaryGPPSchro}
\end{equation}

\noindent where
\begin{equation}
\label{a}
L^2=-\left\lbrack\frac{1}{\sin\theta}\frac{\partial}{\partial\theta}\left (\sin\theta\frac{\partial}{\partial\theta}\right )+\frac{1}{\sin^2\theta}\frac{\partial^2}{\partial\phi^2}\right\rbrack
\end{equation}

\noindent is the squared angular momentum operator and $j$ labels the eigen-state $\psi_j$ with eigen-energy $E_j$. To solve this equation, we assume a separation of variables for $\psi_j := \psi_{n\ell m}(r,\theta,\phi) = R_{n\ell}(r) Y_l^m(\theta,\phi)$, where $Y_l^m(\theta,\phi)$ are the spherical harmonics and $R_{n\ell}$ is expressed as $R_{n\ell} := u_{n\ell}/r$, with $u_{n\ell}$ satisfying the following radial equation: 

\begin{equation}
-\dfrac{\hbar^2}{2 m_B}\dfrac{d^2 u_{n\ell}}{dr^2} + \left(\dfrac{\hbar^2}{2m_B}\dfrac{\ell(\ell + 1)}{r^2} + m_BV_T(r) \right)u_{n\ell} = E_{nl} u_{nl},
\label{eq:unl}
\end{equation}

\noindent where $n$, $\ell$, and $m$ are ``quantum numbers", and where we have  used the identity $L^2Y_{lm}=l(l+1)Y_{lm}$. We name the wave function $\Psi_0$ as the one that fits the target density, which we express as a linear combination of the eigen-functions $\psi_j$:

\begin{equation}
\Psi_0 = \sum_j a_j \psi_j e^{-iE_j t / \hbar}.
\label{eq:waveexpantion}
\end{equation}

\noindent The density profile $|\Psi_0|^2$ associated with the wave function is given by

\begin{eqnarray}
    |\Psi_0|^2 &=&\left(\sum_{j} a_{j}\psi_{j}e^{-iE_j t / \hbar}\right) \left(\sum_{k} a_{k}^* \psi_{k}^*e^{iE_k t / \hbar}\right) \nonumber\\
    &=& \sum_j |a_{j}|^2 |\psi_{j}|^2 + \sum_{j\neq k} a_j a_k^* \psi_j \psi_k^* e^{{\rm i}(E_k - E_j)t/\hbar }.
    \label{eq:wavedensity}
\end{eqnarray}

\noindent An essential assumption when fitting structure densities in structure formation simulations or multi-core collisions, is that $\rho_T$ is a time-averaged quantity, as well as a spatially averaged quantity along various radial directions. Therefore, we assume that the target density can be decomposed as follows:

\begin{eqnarray}\label{psi0}
    \expval{|\Psi_0|^2}_{T\to\infty} & := & \lim_{T\to\infty} \dfrac{1}{T}\int_0^T |\Psi_0(t,\Vec{x})|^2 dt \nonumber\\
    & = &  \dfrac{1}{4\pi}\sum_{n,\ell} (2\ell +1) |\tilde{a}_{n\ell}|^2  |R_{n\ell}|^2, \label{eq:density time average}
\end{eqnarray}

\noindent where $T$ is the time-window used to calculate \textit{time-averages}. To derive Eq. (\ref{psi0}) we have used the identity $\sum_m  |Y_{lm}(\theta,\phi)|^2=(2l+1)/4\pi$. In this formula, the coefficients are written as $a_{n\ell m} = \tilde{a}_{n\ell} e^{i\Theta_{n\ell m}}$, where $\Theta_{n\ell m}$ are random phases with values between $0$ and $2\pi$. Another alternative is to consider a \textit{spatial-average} over the solid angle $\Omega := [0,\pi]\times[0,2\pi]$ as follows:

\begin{eqnarray}
\expval{|\Psi_0|^2}_{\Omega} & := & \dfrac{1}{4\pi} \int_\Omega |\Psi_0(t,\vec{x})|^2d\Omega \nonumber \\
& = & \dfrac{1}{4\pi}\sum_{n,\ell} (2\ell +1) |\tilde{a}_{n\ell}|^2  |R_{n\ell}|^2.
\label{eq:density spatial average}
\end{eqnarray}

\noindent In this way, temporal and  spatial averages are assumed equal and the target density must satisfy $\rho_T \approx m_B \expval{|\Psi_0|^2}_{T\to\infty} = m_B \expval{|\Psi_0|^2}_{\Omega}$. Then, we can simply write $\rho_T \approx m_B \expval{|\Psi_0|^2}$ referring to either angular or time average. However, it must also hold that $V_T \approx \expval{V}$.

An important aspect of $\Psi_0$ is whether it corresponds to a virialized configuration or not. In order to answer this question, we calculate the quantity $Q_0 = 2K_0 + W_0$, where

\begin{equation}
    K_0 =\frac{\hbar^2}{2m_B}\int |\nabla\Psi_0|^2\, d^3x= -\frac{\hbar^2}{2m_B}\int_D \Psi_0^*\nabla^2\Psi_0 d^3x,
    \label{eq:K0}
\end{equation}

\noindent is the kinetic energy and

\begin{equation}
    W_0 = \frac{m_B}{2} \int_D V_T |\Psi_0|^2 d^3x
    \label{eq:W0}
\end{equation}

\noindent is the gravitational energy. In an ideally virialized configuration $Q_0=0$. This quantity can be written in terms of a spectral decomposition as

\begin{equation}
    Q_0 = \sum_{n,l}(2\ell+1)|\tilde{a}_{n\ell}|^2Q_{n\ell},
    \label{eq:Q0_1}
\end{equation}

\noindent with $Q_{n\ell} = 2K_{n\ell} + W_{n\ell}$, where $K_{n\ell}$ and $W_{n\ell}$ are the matrix elements of the kinetic and potential energies with respect to the supposed basis of the eigenproblem given by

\begin{equation}\label{kn}
    K_{n\ell} = -\frac{\hbar^2}{2m_B}\int R_{n\ell}\left[\frac{d}{dr}\left(r^2\frac{d R_{n\ell}}{dr}\right)-\ell(\ell+1)R_{n\ell}\right] \,dr,
\end{equation}

\noindent and 

\begin{equation}\label{wn}
    W_{n\ell} = \frac{m_B}{2}\int V_T R_{n\ell}^2 r^2 \,dr.
\end{equation}

\noindent On the other hand, from Eqs. (\ref{eq:unl}), (\ref{kn}) and (\ref{wn}) we obtain the identity $K_0+2W_0=E_0$ with

\begin{equation}
\label{ber1}
E_0=\frac{1}{m}\int \sum_{nl}  R_{nl}(r)^2 (2l+1) |{\tilde a}_{nl}|^2 E_{nl} r^2\, dr,
\end{equation}

\noindent where $E_{n\ell}$ are the eigenvalues (notice that $E_0$ is an eigenvalue and not the total energy that could be confused with $K_0+W_0$). Therefore, we have $Q_0=2E_0-3W_0$ (in components form $K_{n\ell}+2W_{n\ell}=E_{n\ell}$  and  $Q_{n\ell} = 2E_{n\ell} - 3W_{n\ell}$). The mass reads

\begin{equation}
\label{ber2}
M_0=\int \rho\, d^3x=m_B \int |\Psi_0|^2\, d^3x,
\end{equation}

\noindent hence

\begin{equation}
\label{ber3}
M_0=\int \sum_{nl}  R_{nl}(r)^2 (2l+1) |{\tilde a}_{nl}|^2 r^2\, dr.
\end{equation}

\noindent We numerically verify that, in general, the individual terms $Q_{n\ell}$ are different from zero and can have different sign for different values of $n$ and $\ell$. That is, each mode of superposition is not virialized, however we  find a superposition such that $Q_0 \approx 0$.

The construction of $\Psi_0$ reduces to the calculation of the coefficients $\tilde{a}_{n\ell}$ of the expansion for the target density in equation (\ref{eq:density time average}) or equivalently (\ref{eq:density spatial average}) and a specific way in which the constraint $Q_ 0\approx 0$ is satisfied. Once these coefficients are determined, it becomes possible to reconstruct a wave function that is consistent with the stationary SP system and at the same time has an average density consistent with a core-halo target density.

The steps to construct the FDM core-halo configuration are summarized as follows:

\begin{enumerate}
\item Start with a given target density $\rho_T$. In order to have a finite integrated mass we follow the recipe in \cite{YavetzLiHui2022} that suggests to modulate the target density with a Gaussian $e^{-r^2/(2 r_{0}^2)}$, having $r_{0}$ as the value for which $\frac{\rho_T(0)}{\rho_T(r_0)} \sim  10^{3}$.

\item Solve  Poisson equation (\ref{eq:stationaryGPPa}-\ref{eq:stationaryGPP}) for such density in the domain $r\in[0,2r_{0}]$, as also suggested in \cite{YavetzLiHui2022}.

\item Use the resulting gravitational potential $V_T$ to solve equation \eqref{eq:unl} for all combinations of $n$ and $\ell$ to be considered.

\item Then find the coefficients $a_{j}$ that minimize an error function between $\rho:=m_B \expval{|\Psi|^2}$ and $\rho_T$.
\end{enumerate}

In the following subsection, we elaborate on the ingredients of step 4.

\subsection{Description of the fitting method}

The expansion of the wave function (\ref{eq:waveexpantion}) can have a large number of terms, and they have to be such that $\rho - \rho_T$, or a norm of it, is minimum. Notice that this problem is reduced to locating the minimum of a function (a given norm of the error) that depends on a large number of parameters (the coefficients of expansion (\ref{eq:waveexpantion})). The use of a Genetic Algorithm (GA) for the solution of this type of problems is an option, on the one hand because the concept of DNA can use a large number of genes that we identify with the coefficients of the expansion, and on the other hand the mutation of DNA allows the possibility to exit from local minima. 

A GA is inspired by the theory of evolution, which assumes the existence of a population of individuals characterized by their DNA composed of a chain of genes, which determine their fitness for survival in an environment. The best-adapted individuals, whose adaptation is measured in terms of a fitness function, have a greater chance of surviving and consequently a greater chance of reproducing, passing their genes on to their offspring in a new generation. Offspring results from a combination of their parents' genes, while a speed up in evolution is due to mutation. This process over many generations leads to a population much better adapted than the initial one. 

We adapt this method to solve our minimization problem as follows. We start with a population of individuals, each one with random coefficients in the expansion (\ref{eq:waveexpantion}), that may evolve towards a population of individuals that approach the condition that a norm of $\rho - \rho_T$ is small. For this purpose we define a population of individuals, each one with genes given by the sequence of coefficients $a_j$ of the expansion (\ref{eq:waveexpantion}) and DNA given by the set of coefficients \(\tilde{a}_{n\ell}\). The maximum number of genes considered is \(N_\text{DNA} = n_{\text{max}} \ell_{\text{max}}\), where the quantum numbers take on the values \(n = 1,2,\ldots,n_{\text{max}}\) and \(\ell = 0,1,\ldots,\ell_{\text{max}}-1\). We define the fitness of an individual as a decreasing function of the difference between $\rho$ and $\rho_T$, which is bigger for better-adapted individuals and is near a virialized state such that $Q_0 \approx 0$. We thus define the fitness function as

\begin{eqnarray}
&& \eta = \dfrac{1}{1+|Q_0|}\left[\frac{1}{r_\text{max}}\int_0^{r_{\text{max}}}\frac{(\rho_T - \rho)^{2}}{\rho_T} \, dr\right]^{-1},
\end{eqnarray}

\noindent where the term $1+|Q_0|$ is not significant when $|Q_0| < 1$, but when $|Q_0| > 1$, the value of the fitness function decreases for profiles far from a virialized state. Finally, $r_\text{max} = 2 r_{0}$ is the upper boundary of the numerical domain where the eigenvalue problem (\ref{eq:stationaryGPPSchro}) is solved. 

The operation of the GA is based on the random generation of an initial population of $N_{\text{org}}$ organisms. 
We calculate the fitness function $\eta$ of all indivuduals, and choose the $k$ most fitted organism. Following an elitist approach, these selected individuals prevail through the next generation. From these $k$ organisms, $N_{\text{cross}}$ are randomly chosen to cross-over and produce children for the next generation; in a biological context, one would typically choose $N_{\text{cross}} = 2$, but this is not a limitation in a GA and $N_{\text{cross}} = 5$ worked better. These selected parents will randomly share their genetic material, namely the coefficients of the expansion, to create a new individual. This process is repeated $N_{\text{org}} - k$ times until the initial population size $N_{\text{org}}$ is completed again.

The organisms in the new generation can potentially adapt more effectively with a mutation process that works as follows. We generate a new random number $\beta_{n\ell}$ ranging from 0 to 1, representing the likelihood that the gene $\tilde{a}_{n\ell}$ undergoes a mutation. Each gene has its own probability of change. Subsequently, a new random number $\gamma_{n\ell}$ is generated, and the mutation occurs if $\gamma_{n\ell}$ exceeds $\beta_{n\ell}$. In such cases, the coefficient $\tilde{a}_{n\ell}$ is altered to $\alpha \tilde{a}_{n\ell}$, where $\alpha$ is a randomly selected number within the range of -1.5 to 1.5 for all values of $n$ and $\ell$.

Finally, a second type of mutation, known as differential mutation is applied. This mutation involves selecting the \textit{i}-th organism with DNA defined by the coefficients $\tilde{a}_{n\ell}^{(i)}$ along with a fitness $\eta^{(i)}$. Subsequently, two other organisms with DNA $\tilde{a}_{n\ell}^{(1)}$ and $\tilde{a}_{n\ell}^{(2)}$ are randomly selected. A new organism is then created by linearly combining these coefficients as $\tilde{a}_{n\ell}^{(new,i)} = \tilde{a}_{n\ell}^{(i)} + \delta(\tilde{a}_{n\ell}^{(1)} - \tilde{a}_{n\ell}^{(2)})$, where $\delta$ is a number between 0 and 1, with fitness $\eta^{(new,i)}$. If it happens that $\eta^{(\text{new},i)}>\eta^{(i)}$, the \textit{i}-th organism is replaced by the new organism. This process is repeated for $i=1,2,\ldots,N_\text{org}$.

Notice that the fitness function is a norm of the error between the density of the multipolar expansion and the target density. Considering the randomness in various stages of the GA it could well happen that different sets of coefficients of the expansion, or equivalently individuals with different DNA, may have similar values of $\eta$. In this sense, the expansion of the profile can be degenerate.

Now, our goal is to tune the galactic dark matter densities. Inspired by \cite{PhysRevD.97.103523} and \cite{YavetzLiHui2022} we use a certain type of target density profile which we discuss below.

\subsection{Models for target density}

\textit{Core-NFW model.} Simulations of binary systems, multicore mergers, and more complex scenarios like structure formation simulations reveal that the time-spatial averages of the formed structures exhibit a spherical profile with a soliton core at the center. This core, with density similar to that of the ground state of the SP system (see e.g. \citet{GuzmanUrena2004}), is modeled with the empirical profile \citep{Schive:2014hza,Schive:2014dra}:

\begin{eqnarray}
&& \rho_{\text{core}}(r) = \rho_{c} \left[1+0.091\left(\frac{r}{r_c}\right)^{2} \right]^{-8},
\label{eq:coreprofile}
\end{eqnarray}

\noindent where we can find the relation between the central density $\rho_c$ and the core radius $r_c$ from the numerical solution of the ground state using the boundary condition $\psi(r=0)=1$ in units where $\hbar = m_B = 4\pi G = 1$. If we fix $\rho_c = 1$, we can find that $r_c \approx 1.30569 \pm 0.000113$. Using the $\lambda$-scaling relation of the SP system \cite{GuzmanUrena2004}, it is found that $\rho_c \approx (1.30569/r_c)^4$ for an arbitrary value of the core radius $r_c$. With this, we can translate the central density to physical units as

\begin{equation}
\begin{matrix}
 \rho_c & = & \dfrac{\hbar^2}{4\pi G m_B^2}\left(\dfrac{1.30569}{r_c}\right)^4 \\ 
 & \approx & 1.983\times10^{7} \left( \dfrac{\textrm{kpc}^4}{m_{22}^2r_c^4}\right) M_\odot,
\end{matrix}
\end{equation}

\noindent where $m_{22}$ is defined as $m_{22} = m_B \times 10^{-22}\mathrm{eV}^{-1}$, and the units for $r_c$ are $[r_c] = \mathrm{kpc}$.

Outside of this core, there is an envelope region that can be approximated by the NFW profile \citep{Schive:2014dra}:

\begin{eqnarray}
&& \rho_{\text{NFW}}(r) = \frac{\rho_{s}}{\frac{r}{r_s}\left(1+\frac{r}{r_s}\right)^2},
\label{eq:NFW}
\end{eqnarray}

\noindent where $\rho_{s}$ and $r_s$ are halo parameters. The complete profile of the structure takes the form \citep{Marsh-Pop:2015}:

\begin{eqnarray}
    \rho_{CN}(r) & = & \rho_{\text{sol}}(r)\Theta(r-r_t) + \rho_{\text{NFW}}(r)\Theta(r_t-r).
    \label{eq:targetDensity coreNFW}
\end{eqnarray}

\noindent In this equation, we assume continuity, which fixes one of the two halo parameters with the relation $\rho_s = \rho_{\text{sol}}(r_t)\frac{r_t}{r_s}\left(1+\frac{r_t}{r_s}\right)^2$.

\textit{Core-PISO model.} It is well-known that a soliton nucleus forms within the halo in the FDM model since the ground state is an attractor of the SP system. However, in the envelope region, it is possible to discuss what may be the best approximation for the average profile of the envelope. One of the alternative proposals to the NFW model is the Pseudo-Isothermal profile, which is written as

\begin{eqnarray}
&& \rho_{\text{PISO}}(r) = \frac{\rho_{p}}{1+\left(\frac{r}{r_p}\right)^2},
\label{eq:PISO}
\end{eqnarray}

\noindent in this case, $\rho_{p}$ and $r_p$ are halo parameters. The complete profile of the structure takes the form:

\begin{eqnarray}
    \rho_{CP}(r) & = & \rho_{\text{sol}}(r)\Theta(r-r_t) + \rho_{\text{PISO}}(r)\Theta(r_t-r),
    \label{eq:targetDensity corePISO}
\end{eqnarray}

\noindent similar to the core-NFW model, in which we assume continuity in the density. In this case the envelope parameters can be related as:

\begin{equation}
    \rho_p = \rho_{sol}(r_t)\left[1+\left(\frac{r_t}{r_p}\right)^2\right].
\end{equation}

\noindent which reduces the number of fitting parameters.

\subsection{Fitting of LSB galaxies} 

LSB galaxies are dominated by dark matter, thus we assume that we can fit their rotation curves with the core-NFW or core-PISO profiles. The independent parameters of a core-NFW profile are $r_c$, $r_t$, and $r_s$, and for the core-PISO profile, they are $r_c$, $r_t$, and $r_p$. We use the same strategy presented in \citet{Bernal:2017oih} to obtain the appropriate values for observational data in \citep{deBlok:2001,McGaugh:2001}. We additionally find the radius $r_0$ of the resulting configurations. Table~\ref{tab:LSB-comb} provides the best-fit free parameters for these galaxies.

Now, according to \cite{Kormendy:2004se,Spano:2007,Donato:2009ab}, the halo surface density is nearly constant and independent of the galaxy luminosity, with value $\Sigma_0 = \rho_0 r_0 = 140^{+80}_{-30} \ M_\odot \text{pc}^{-2}$, where $\rho_0$ and $r_0$ the halo central density and core radius \citep{Donato:2009ab}. We include in Table~\ref{tab:LSB-comb} the corresponding surface densities for both the core-NFW and core-PISO profiles, as defined in \citet{Chavanis2022vziNEX}, for the soliton FDM configurations: $\Sigma_0 = \rho(r_t) r_h$, with $r_h$ the radius where the density is $\rho(r_h) = \rho(r_t)/4$, for $r_t$ the transition radius outside the soliton region. As discussed in Appendix L of  \cite{Chavanis2022vziNEX}, the constant $\Sigma_0$ observational value is not consistent with the soliton, it decreases like $1/r^3$ as the size of the soliton increases. This suggests (see Sec. VII of \cite{Chavanis2022vziNEX}) to define $\Sigma_0$ with the density at the interface between the soliton and the NFW envelope, at the transition radius $r_t$.

As seen in Table~\ref{tab:LSB-comb}, from the small sample we are studying, the results do not coincide with the observational value, except for ESO4880049 with the core-PISO profile. For the core-PISO profile, the results are closer to the value obtained in \citep{Donato:2009ab}. The discrepancy may arise from the density profile assumed to model their huge sample of galaxies, a Burkert profile. In our case, the core-PISO profile decays slowly and is closer to the Burkert profile. We would need to simulate a large sample of galaxies with a profile closer to Burkert's to conclude if our results are in agreement with a universal surface density of dark matter.

\begin{table}
\begin{tabular}{lccccc}
\hline
&& core-NFW &&& \\
\hline
Galaxy & $r_c$ & $r_t$ & $r_s$ & $r_0$ & $\Sigma_0$ \\
& (kpc) & (kpc) & (kpc) & (kpc) & ($M_\odot \text{pc}^{-2}$) \\
\hline
ESO4880049 & 2.157 & 1.102 & 15.25 & 52.77 & 243 \\
UGC11616 & 1.860 & 1.676 & 7.434 & 38.04 & 386 \\
F730V1 & 1.867 & 1.841 & 8.118 & 40.01 & 370 \\
\hline
\\
\hline
&& core-PISO &&& \\
\hline
Galaxy & $r_c$ & $r_t$ & $r_p$ & $r_0$ & $\Sigma_0$ \\
  & (kpc) & (kpc) & (kpc) & (kpc) & ($M_\odot \text{pc}^{-2}$) \\
\hline
ESO4880049 & 2.269 & 3.260 & 2.631 & 66.49 & 155 \\
UGC11616 & 1.850 & 2.474 & 1.792 & 52.82 & 294 \\
F730V1 & 1.869 & 2.082 & 1.625 &  54.41 & 345 \\
\hline
\end{tabular}
\caption{Best-fit parameters for three LSB galaxies using the core-NFW and core-PISO density profiles, obtained by fixing the boson mass $m_B = 10^{-23}\mathrm{eV}$. }
\label{tab:LSB-comb}
\end{table}

As described earlier, there is no consensus on the correct mass of the boson, and in our study we use a boson mass $m_B = 10^{-23}$ eV because it is near the upper bound, and allows the profiles to be adjusted for the LSB galaxies in our analysis. This mass value is on the boundary with the cosmological constraints found in \citep{Bozek:2014} from the galaxy UV-luminosity function and reionization observations, and in \citep{Sarkar:2016} from Lyman-$\alpha$ observations, $m_B > 10^{-23}$ eV. It also falls within the constraints provided by \citep{Hlozek:2014}, from CMB and galaxy clustering data, $m_B > 10^{-24}$ eV. 

Now, concerning the fitting method, the parameters parameters of the GA are a population of $N_\text{org} = 200$ organisms, each having $n_\text{max} = l_\text{max} = 41$. This implies that the DNA of each organism consists of $N_\text{DNA} = 1681$ genes, resulting in a total of approximately $10^{5}$ coefficients $a_{nlm}$, a similar number as in \cite{YavetzLiHui2022}. During reproduction, $N_\text{cross} = 5$ organisms contribute to creating a new organism, selected from a pool of $k = 100$ parents. Additionally, for the differential mutation, we set $\delta = 0.1$. These parameters have proven effective in identifying organisms with a fitness $\eta \approx 10^{5}$, or equivalently, a proportional $\chi^2$ error $1/\eta \approx 10^{-5}$ within the initial 1000 generations, and in general a virialization factor in the range $|Q_0|<10^{-5}$ in code units.

\begin{figure}
\centering
\includegraphics[width=8.0cm]{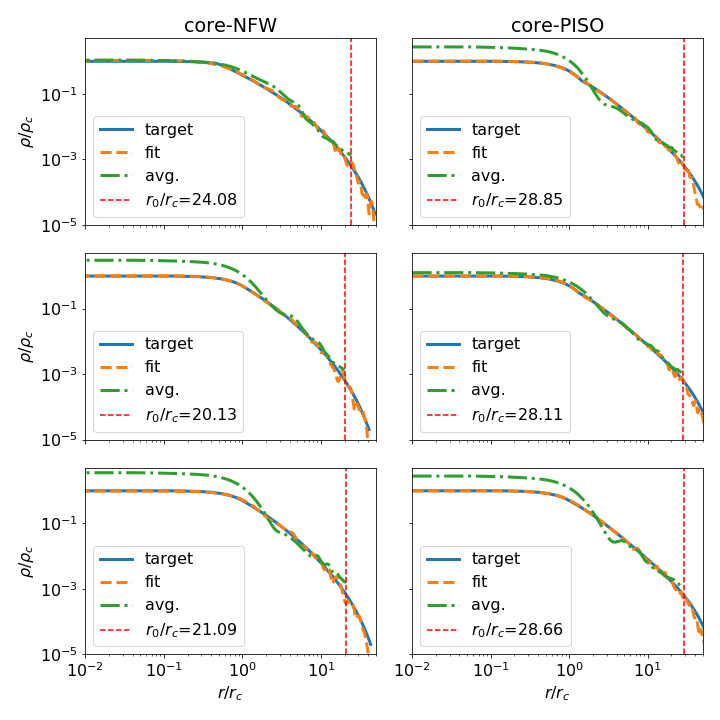}
\caption{Target, fitted, and average density profiles using the GA. The first, second, and third rows correspond to the galaxies ESO4880049, UGC11616, and F730V1, respectively. The vertical red line indicates the radius $r_0$ used in the weighted exponential function for the target density. The first column corresponds to the core-NFW profile, and the second column corresponds to the core-PISO profile. The continuous blue line represents the target density, while the orange dotted line illustrates the fitted density obtained in the 1000th generation. Finally, the green dotted line represents the  average in time of the spatial averaged density of snapshots during 7Gyr of evolution that we describe below; notice the that the evolution clearly distorts the density profile. }
\label{fig:datarhoGA}
\end{figure}

Using these parameters, we determined the suitable coefficients for each of the considered galaxies in Table \ref{tab:LSB-comb}. The results appear in Figure \ref{fig:datarhoGA}, which illustrate how the GA is able to construct multimode configurations that approximate the target density within the region $r < r_0$. Beyond this radius, the adjustment becomes more challenging, as seen after the a vertical red dotted line.

\section{Evolution of the galactic profiles}
\label{sec:stability}

We investigate the evolution of the core-halo profiles described in the previous section by evolving the wave function with the fully time-dependent SP system (\ref{eq:Gross-Pitaevskii}-\ref{eq:Poisson}), for which we use the code CAFE \cite{Alvarez_Rios_2022,periodicas}. In order to prevent the wave function from decaying into an isolated solitonic profile as suggested in \citet{GuzmanUrena2006} and \citet{BernalGuzman2006b}, and later confirmed with CAFE in \cite{periodicas}, we implemented periodic boundary conditions that guarantee the persistence of a core surrounded by an envelope, as well as the constancy of mass and total energy.

As initial conditions, we inject the wave function (\ref{eq:waveexpantion}) at time $t=0$, $\Psi(0,\vec{x})=\Psi_0$ centered in the 3D cubic box $D=[-r_{0},r_{0}]^3$. It is worth noticing that when the coefficients are fixed, the wave function can possess an overall momentum different from zero, calculated as $\vec{p}_0 = -i\hbar \sum_{j,k} a_k^* a_j \int_D \psi_k^* \nabla \psi_jd^3x$. Then, we correct the initial wave function to be $\Psi(0,\vec{x}) = \Psi_0 e^{-i \vec{p}0 \cdot \vec{x} / M}$, where $M :=\int_D \rho d^3x$ represents the total mass in the domain. This choice ensures that the initial wave function has zero total linear momentum, and the evolution remains with the core nearly at the center of the domain. The domain was discretized with a spatial resolution of $\Delta = r_{0}/128$ along the three spatial directions. To capture the temporal dynamics, a time resolution satisfying $\Delta t / \Delta <0.25$ in code units was employed, and the evolution was carried out over a time-window of $2$ Gyrs.

\begin{figure}
\centering
\includegraphics[width=8.0cm]{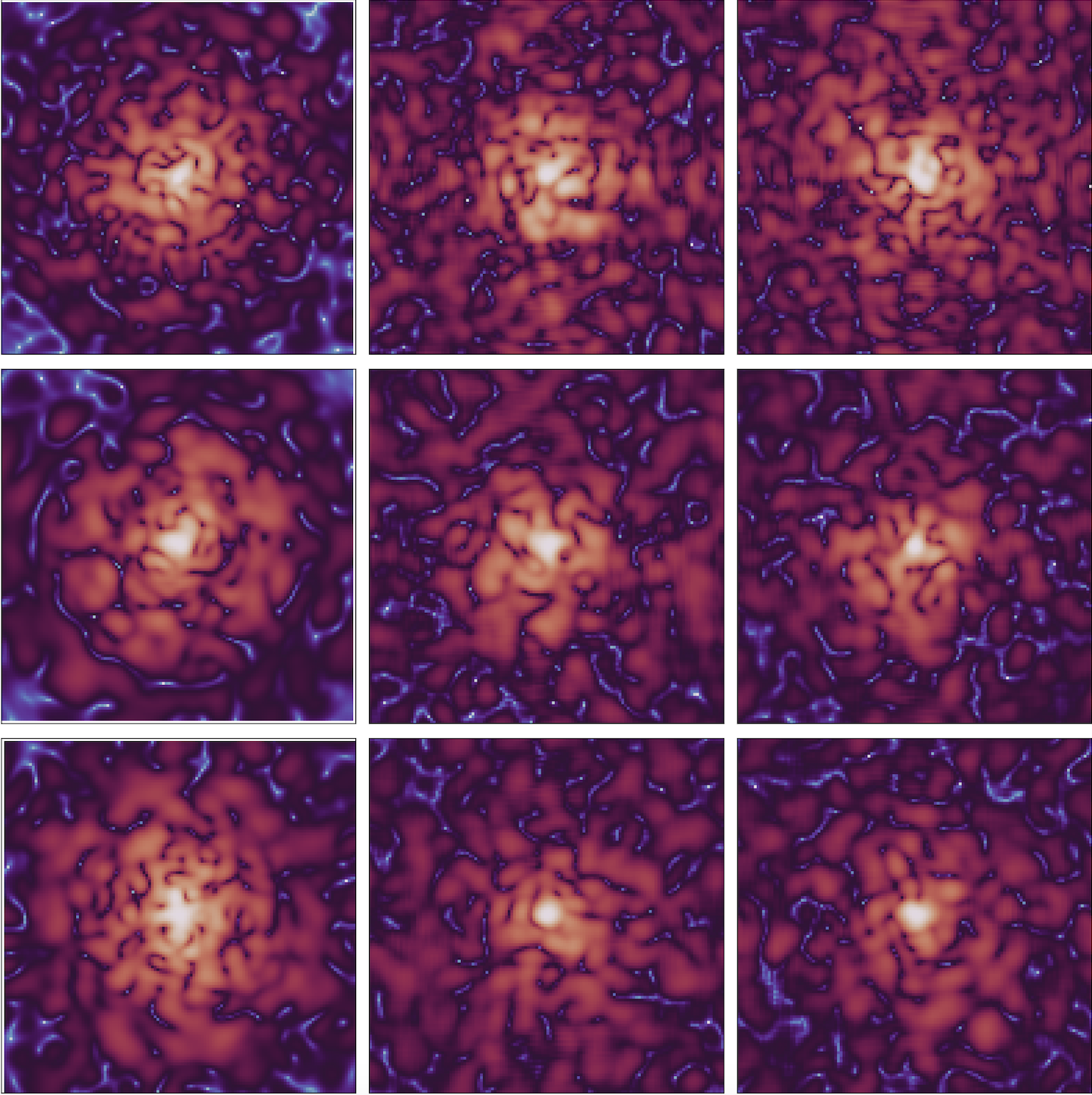}
\includegraphics[width=2.625cm]{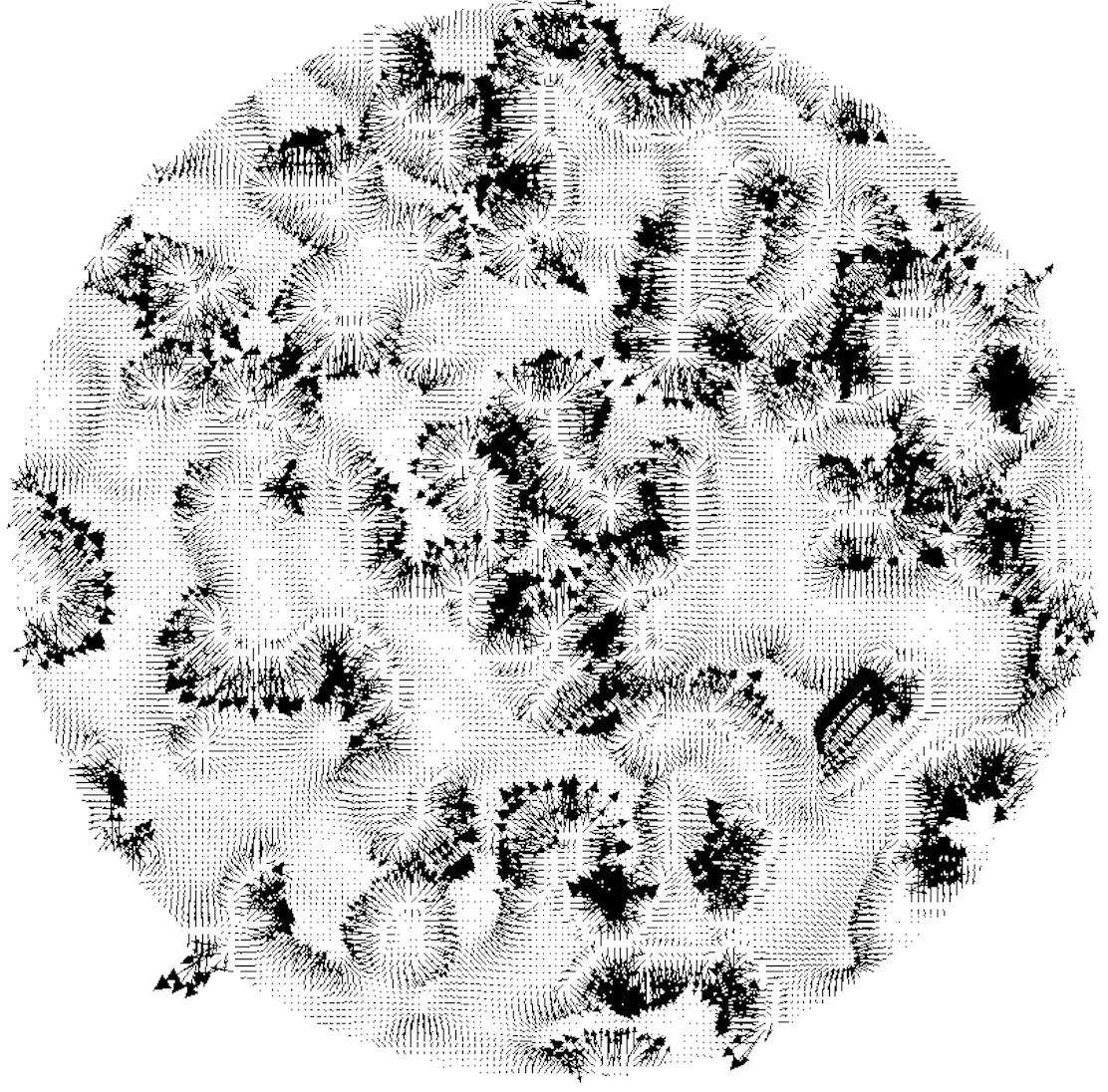}
\includegraphics[width=2.625cm]{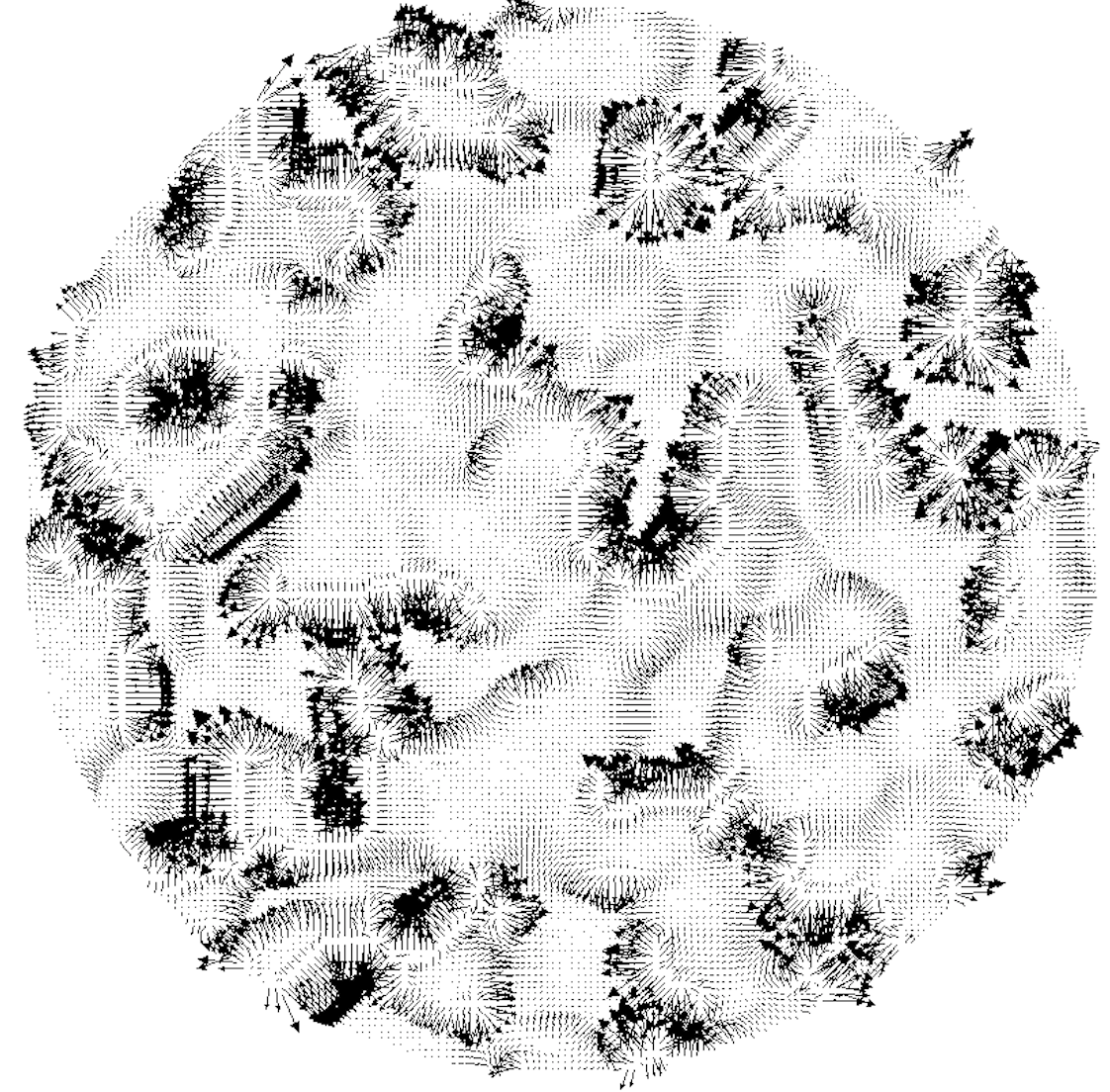}
\includegraphics[width=2.625cm]{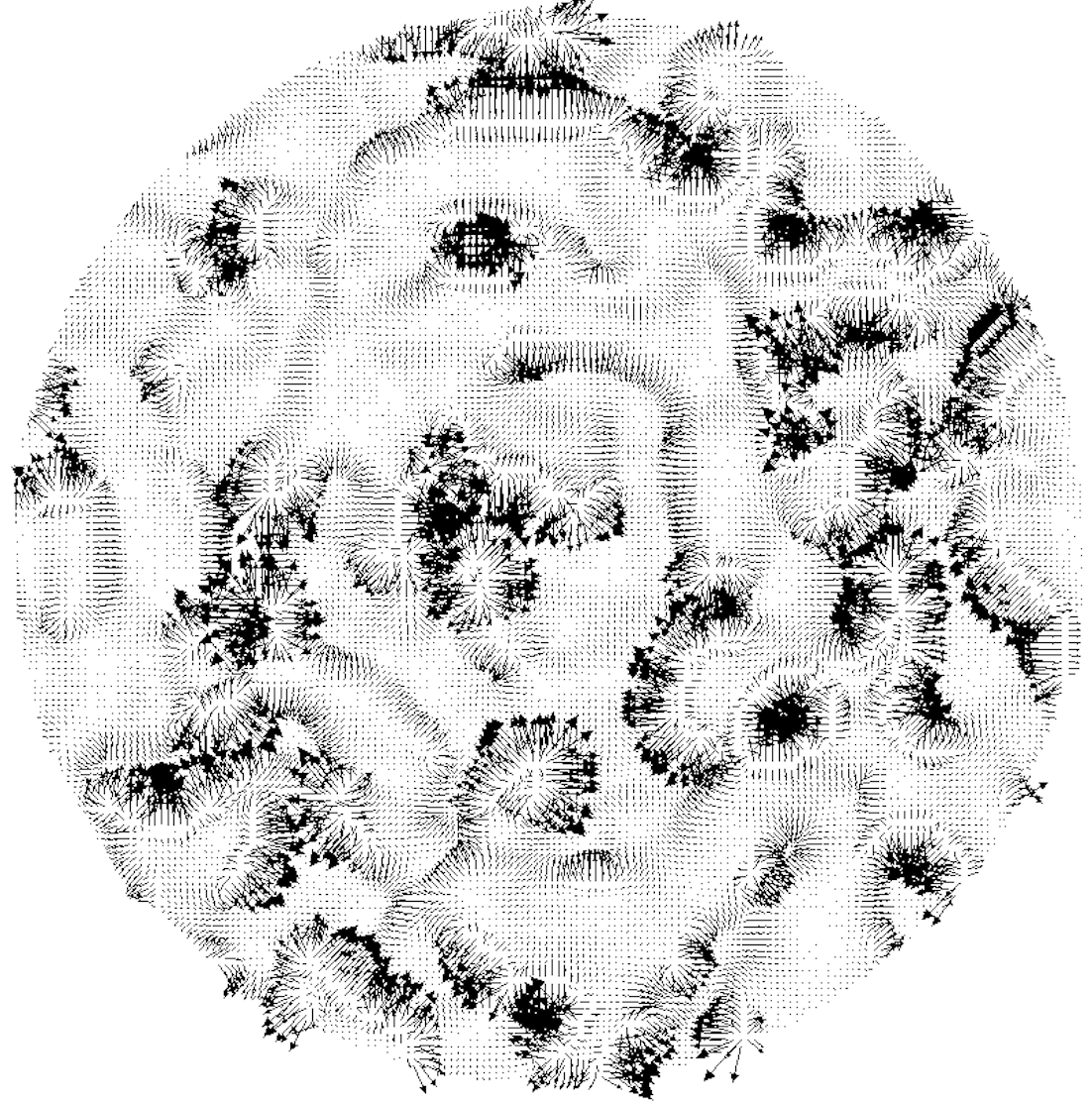}
\caption{The first, second, and third rows show snapshots of the density in the plane $z=0$, for galaxies ESO4880049, UGC11616, and F730V1, respectively, assuming the core-NFW target profile. The first column corresponds to initial time, while the second and third columns correspond to times $t = 1$ and $2$ Gyr, respectively. The fourth row shows snapshots at 0, 1 and 2Gyr of the velocity field for the galaxy F730V1, for illustration.}
\label{fig:densityEvolutionCoreNFW}
\end{figure}

\begin{figure}
\centering
\includegraphics[width=8.0cm]{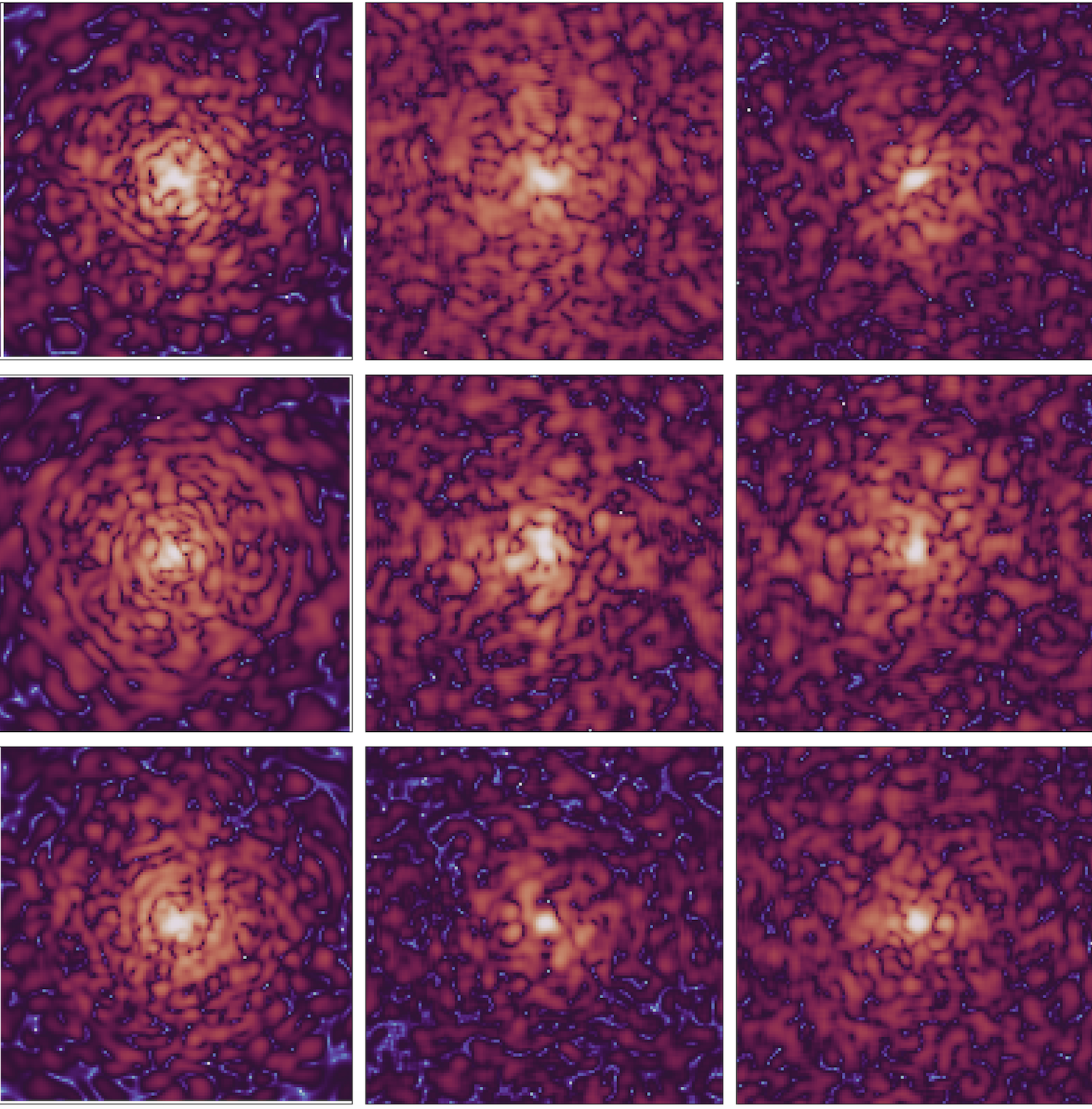}
\includegraphics[width=2.625cm]{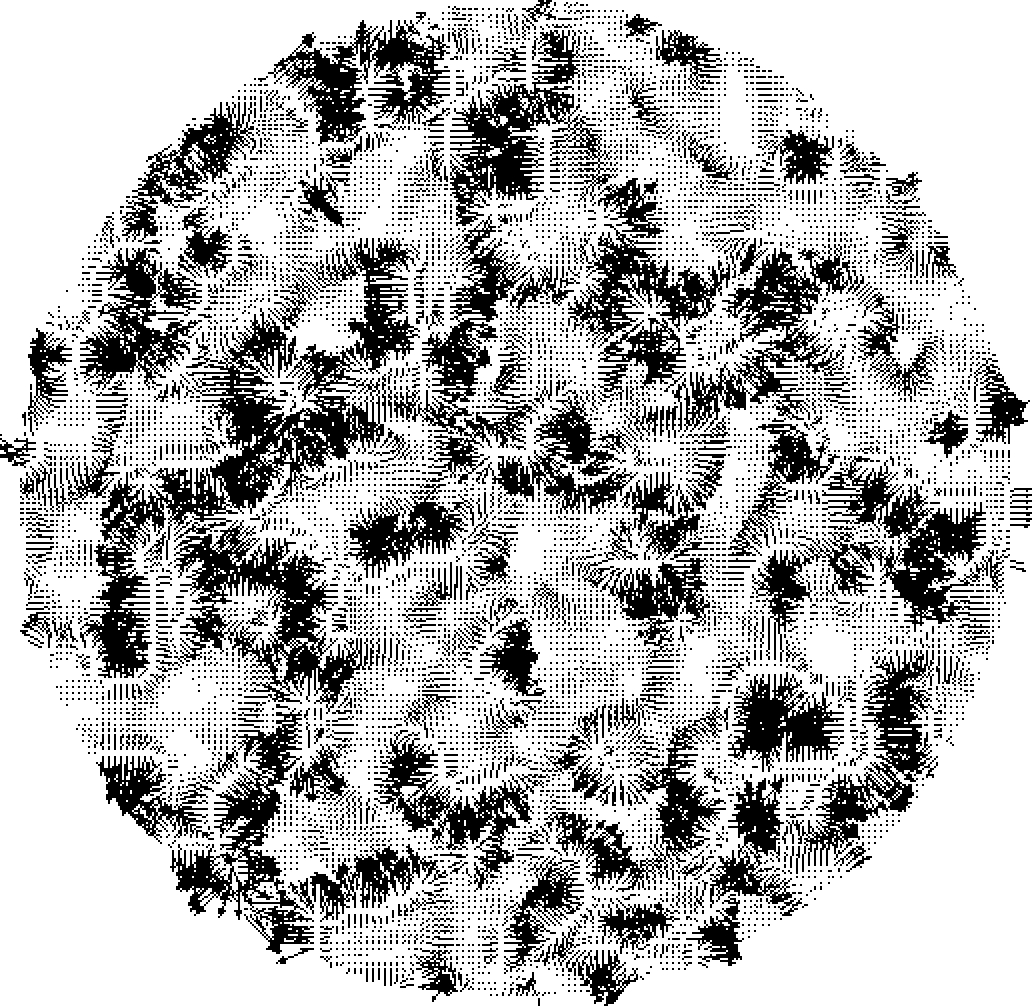}
\includegraphics[width=2.625cm]{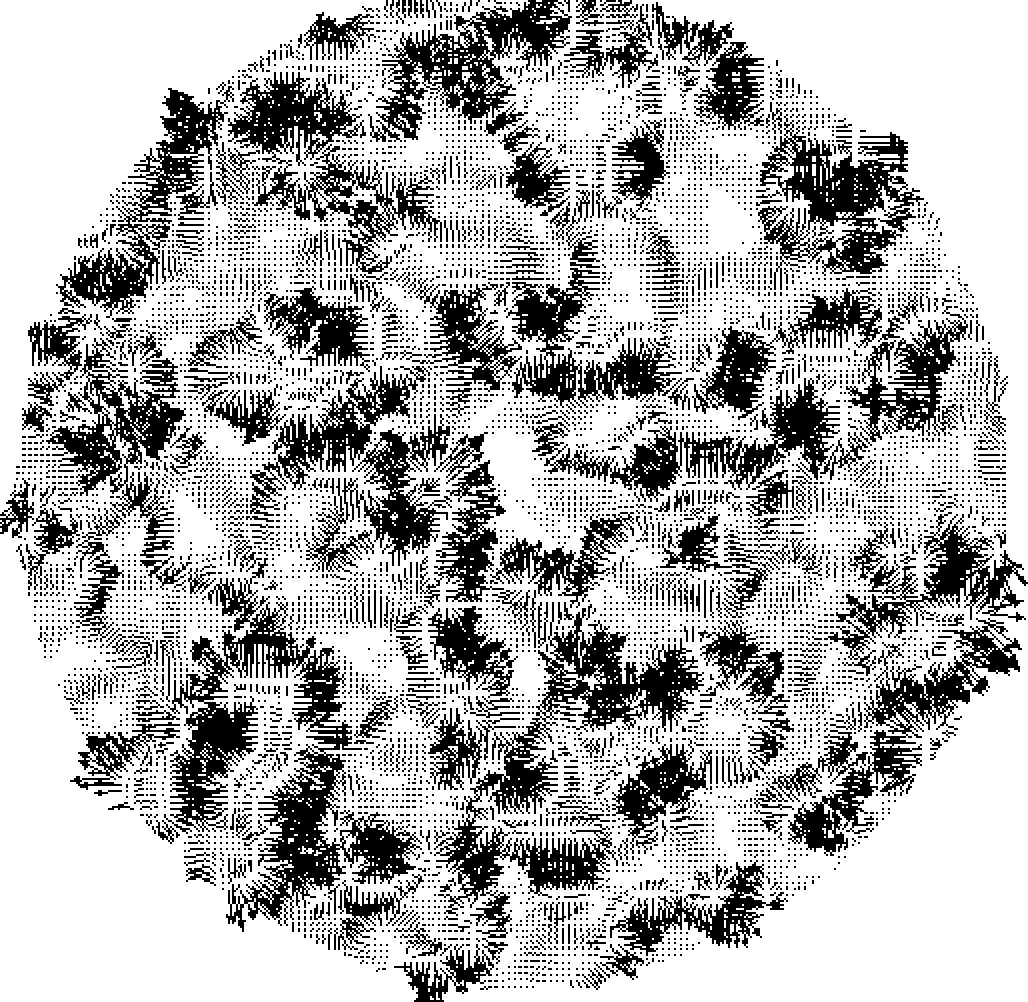}
\includegraphics[width=2.625cm]{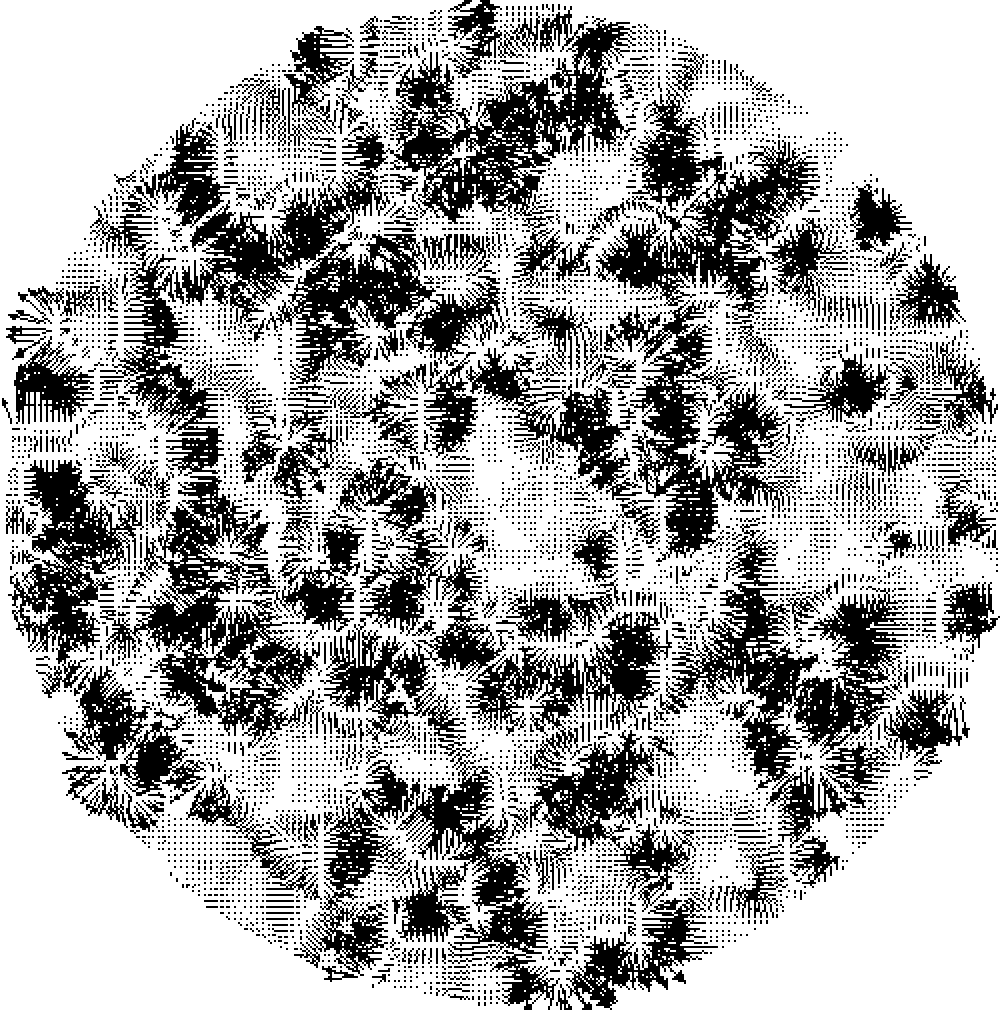}
\caption{The first, second, and third rows show snapshots of the density in the plane $z=0$, corresponding to the galaxies ESO4880049, UGC11616, and F730V1, respectively, assuming the core-PISO target profile. The first column corresponds to initial time, while the second and third columns correspond to times $t = 1$ and $2$ Gyr, respectively. The fourth row shows snapshots at 0, 1 and 2Gyr of the velocity field for the galaxy F730V1, for illustration.}
\label{fig:densityEvolutionCorePISO}
\end{figure}

The evolution of each galaxy is depicted through snapshots of the density and velocity vector field in the $z=0$ plane at times $t=0$, $1$, and $2$ Gyr in Figures \ref{fig:densityEvolutionCoreNFW} and \ref{fig:densityEvolutionCorePISO}. 
These simulations use the initial conditions with core-NFW and core-PISO target density profiles, respectively.

\begin{figure}
\centering
\includegraphics[width=8.0cm]{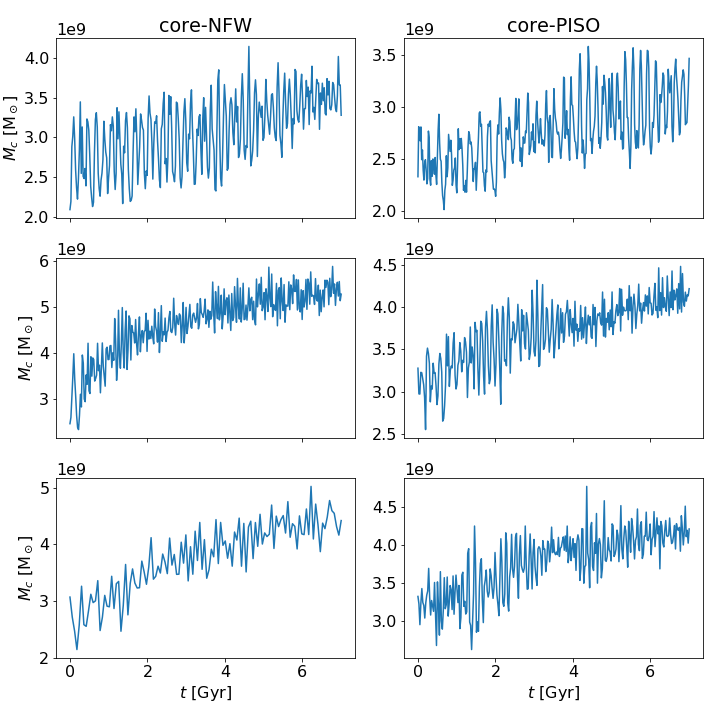}
\caption{Evolution of $M_c$ as function of time during 7Gyr. The first, second, and third rows correspond to the galaxies ESO4880049, UGC11616, and F730V1, respectively. The first column corresponds to the core-NFW profile, and the second column corresponds to the core-PISO profile. The plots indicate that the core accretes mass in the long term, which implies a time-dependent concentration of matter. This process is interpreted as the accretion from the granular envelope. This mass growth has been shown to happen after core saturation after relaxation \cite{ChengNiemeyer2021} and seems to be universal in core-halo structures that evolve \cite{LevkovTkachev2018}.}
\label{fig:McAccretion}
\end{figure}

It is evident that even though the configurations are initially near a virialized state, they evolve and in fact the configurations do not remain stationary and not even in average, instead they develop some dynamics. In order to understand better the evolution of the whole configuration, we look into the time dependence of the core mass for each of the galaxies of the sample.  The core mass $M_c$ is the integral of the density (\ref{eq:coreprofile}) from the origin until $r_c$ and its value as function of time is shown in Fig. \ref{fig:McAccretion} for six of the configurations during 7Gyr. Notice that the core mass oscillates with an overall growing trend that can be understood as the accretion of matter from the granular envelope, indicating that the growth mass is due to collisional effects \cite{Bar-or,Chavanis2020}, interpreted as condensation in the kinetic regime \cite{LevkovTkachev2018} or wave condensation \cite{EggenmeierNiemeywr2019}. This slow, but never ending core mass growth, has been shown to happen after the saturation time \cite{ChengNiemeyer2021}. This core mass growth seems inevitable and the reason why possibly any configuration with granular structure will lead to evolution and core growth. As a result, the dynamics is influenced and the average density in the evolution deviates from the averages of the initial data, an effect also described in \cite{YavetzLiHui2022}.

\begin{figure}
\centering
\includegraphics[width=8.0cm]{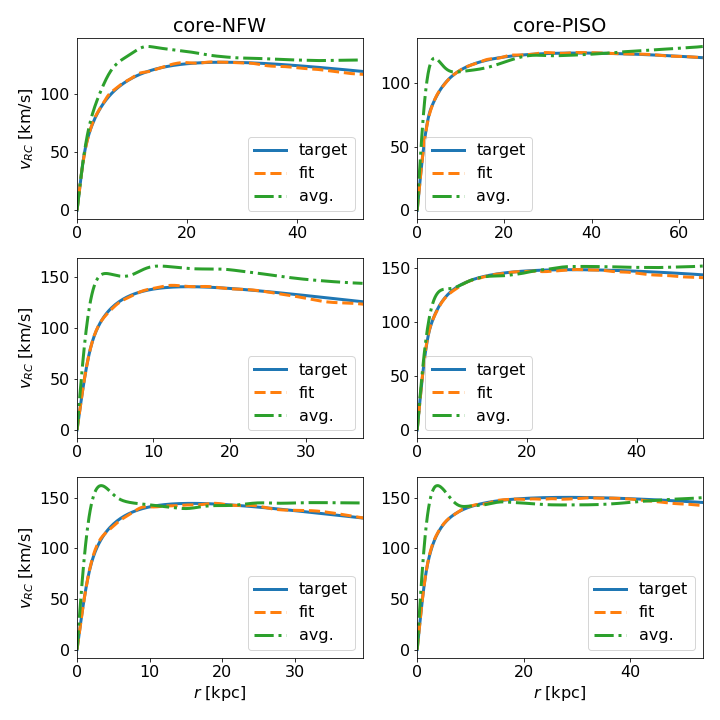}
\caption{Target, fitted, and average rotation curve of three galaxies of the sample. The first, second, and third rows correspond to ESO4880049, UGC11616, and F730V1, respectively. The first column corresponds to the core-NFW profile, and the second column corresponds to the core-PISO profile. The continuous blue line represents the rotation curve associated with the target density, while the orange dotted line illustrates the fitted density obtained in the 1000th generation. The green dotted line represents the time-average of the spatial averages during 7Gyr of evolution.}
\label{fig:datavRC}
\end{figure}

The implication is that the redistribution of density will also distort the rotation curve. We illustrate this difference by calculating the spatio-temporal density average $\expval{\rho}$, which is now only a function of the radial coordinate. Once the average density is obtained, we compute the radial rotation curve as $v_{RC} = \sqrt{G m(r)/ r}$, where $m(r)$ is the mass density integrated until radius $r$. For each halo, the results are presented in Figure \ref{fig:datavRC}. The discrepancy between the initial and the evolved configuration is noticeable. The concentration of matter in the core region definitely changes the RC with a characteristic peak of a concentrated mass.

\section{Conclusions}
\label{sec:conclusions}

We present a method to construct FDM halos, with multimode expansions characterized by a core-halo profile associated to observational rotation curves. The target density profiles were of type core plus an envelope with NFW and Pseudo-Isothermal profiles that fit rotation curves of a sample of LSB galaxies. While the core is dominated by the first term in an expansion in spherical harmonics, the envelope contains the expected granular structure. When averaged on the solid angle, the density profile approaches the target density that fits rotation curves.

Even though the constructed configurations are nearly virialized, it seems unavoidable the evolution of the configuration that degrades the quality of the RCs fittings. We then evolved these configurations during 7Gyr and measured the core mass as function of time, we found the generic result, namely, the core permanently accretes matter from the granular envelope, an effect already measured after the formation and saturation time of cores \cite{ChengNiemeyer2021}, with a slow but permanent growth  that goes as $t^{1/8}$.  The core-halos are stable in the ``collisionless'' regime but they evolve due to ``collisions'' (granularities) on a secular timescale. Note that this is not an instability but a natural secular dynamical evolution, accompanied by a condensation process and the growth of the soliton. This is in agreement with kinetic theory as described in \cite{Bar-or,Chavanis2020}.

A direct implication of the core growth is that matter concentrates near the center of the galaxy and the rotation curve develops a characteristic peak at a small radius, observed in some galaxies (e.g. in \cite{Bernal:2017oih}). A lesson from our analysis is that no matters how well RCs are fitted with a core surrounded a granular envelope, and how virialized to model is, FDM configurations will evolve and get distorted by the core accretion.

A considerable enhancement to this analysis would be the contribution of luminous matter during the evolution, which if gravitationally coupled to FDM and would influence the dynamics of the whole structure.


\section*{Acknowledgments}
Iv\'an \'Alvarez receives support within the CONAHCyT graduate scholarship program under the CVU 967478. FSG is supported by grants CIC-UMSNH-4.9. TB and FSG are supported by CONAHCyT Ciencia de Frontera 2019 Grant No. Sinergias/304001. 


\appendix

\section{Connection between the wave description and the kinetic description}

In this Appendix, we discuss the connection between the wave description and the kinetic description. We recall the relation between the wave superposition coefficients $|{\tilde a}_{nl}|^2$ and the  particle distribution function $f(\epsilon)$ following \cite{YavetzLiHui2022}. We then use this relation to recover the classical energy functionals and the classical virial theorem from the quantum ones in the WKB (high energy) limit.

\subsection{Classical kinetic description based on the Vlasov-Poisson equations}

In a classical description (applying for example to stellar systems or to the time-averaged envelope of DM halos) the density is given by
\begin{eqnarray}
\label{a1}
\rho=\int  f\, d{\bf v},
\end{eqnarray}
where $f({\bf r},{\bf v},t)$ is the distribution function for particles of mass $m_B$, i.e., $f({\bf r},{\bf v},t)$ gives the mass density of particles with position ${\bf r}$ and velocity ${\bf v}$ at time $t$. It is normalized such that $\int f\, d{\bf r}d{\bf v}=M_{\rm envelope}$. We assume that the envelope is spherically symmetric with a DF $f=f(\epsilon)$ that is a function of the energy alone. Such a DF determines a steady (virialized) state of the classical Vlasov-Poisson equations. We have introduced the energy per unit mass $\epsilon=E/m_B=v^2/2+V$, where $V$ is the gravitational potential. Eq. (\ref{a1}) can then be rewritten as
\begin{eqnarray}
\label{a2}
\rho=\int_{V(r)}^0 f(\epsilon)4\pi  \sqrt{2(\epsilon-V)} \, d\epsilon.
\end{eqnarray}

In practice it is difficult to predict the DF $f$ of the envelope resulting from the process of violent relaxation. Note that there is no rigorous derivation of the NFW and Burkert profiles so these profiles remain essentially empirical.  Actually, a prediction of the DF may be attempted from the statistical  theory of  Lynden-Bell \cite{lb1967,Chavanis_1996} However, this ``naive'' prediction leads to a DF with an infinite mass so it is necessary to take into account the evaporation of high energy particles to have a more physical model.  This leads to the  fermionic King model \cite{c1998} 
where the ``fermionic'' nature of the DF arises from the specificities of the Vlasov equation in the Lynden-Bell statistical theory. In many cases ``degeneracy'' effects can be neglected leaving us with the classical King model.  The King model determines a sequence of equilibrium states (indexed by the central concentration) ranging from a pure polytrope of  index $n=5/2$ to an isothermal distribution ($n=\infty$) \cite{king1965,clm2015}. It is shown that the King model at the critical point of marginal stability, just before the system undergoes the gravothermal catastrophe, gives a good agreement with the Burkert profile (see the comparison between the different density profiles reported in Fig. 18 of \cite{clm2015} and Fig. 1 of \cite{c2022}. Therefore, the (fermionic) King model may be a relevant model of DM halos that is physically motivated.

\subsection{Quantum wave description based on the Schr\"odinger-Poisson equations in the WKB approximation}

The quantum wave description of DM halos is discussed in the main text. Because of the process of violent relaxation or gravitational cooling, the envelope of DM halos may be viewed as a superposition of excited states with energies $E_{nl}$ and amplitude ${\tilde a}_{nl}$. This is similar to the orbits of particles with energies $\epsilon$ and DF $f(\epsilon)$  in classical systems. By contrast, the core (soliton) of quantum DM halos corresponds to the ground state of the Schr\"odinger-Poisson equations that has no counterpart  in classical systems. Here, we focus on the envelope and we consider sufficiently high energies $E$ so that the WKB approximation can be employed (see \cite{YavetzLiHui2022} for details).

In the WKB approximation (large $E$ limit), the radial function is given by 
\begin{equation}
\label{a3}
R_{nl}(r)=\frac{N_{nl}}{r\sqrt{p(r)}}\sin\left\lbrack \frac{1}{\hbar}\int_{r_1}^r p(r')\, dr'+\frac{\pi}{4}\right\rbrack,
\end{equation}
where
\begin{equation}
\label{a4}
p(r)=\sqrt{2m_B\left (E_{nl}-\frac{l(l+1)\hbar^2}{2mr^2}-m_B V\right )}
\end{equation}
is the classical radial momentum. The normalization condition is chosen such that $\int_{r_1}^{r_2}  R_{nl}(r)^2 r^2\, dr=1$ 
 giving
 
\begin{equation}
\label{a5}
N_{nl}^2=\frac{1}{\int_{r_1}^{r_2} \frac{dr}{2p(r)}},
\end{equation}
where we have approximated the square of the sine as  $1/2$. In the above expressions, $r_1$ and $r_2$ are the turning points where $p$ vanishes.  The energy eigenvalues $E_{nl}$ satisfy the  Bohr-Sommerfeld quantization condition
\begin{equation}
\label{a6}
\int_{r_1}^{r_2} p(r)\, dr=\pi\hbar \left (n+\frac{1}{2}\right ).
\end{equation}
To compute the time-average density of the envelope we approximate the sum over $n$ and $l$ in Eq. (\ref{psi0}) by integrals and write
\begin{eqnarray}
\label{a7}
\rho(r)=\frac{1}{4\pi}\int d\epsilon dl\, \frac{dn}{d\epsilon} R_{nl}(r)^2 (2l+1) |{\tilde a}_{nl}|^2 
\end{eqnarray}
with $\epsilon=E_{nl}/m_B$. The Jacobian $dn/dl$ can be obtained by differentiating the quantization condition from Eq. (\ref{a6}) yielding
\begin{equation}
\label{a8}
\frac{dn}{d\epsilon}=\frac{m^2}{\pi\hbar}\int_{r_1}^r \frac{dr}{p(r)}.
\end{equation}
Using this expression together with the WKB approximation for $R_{nl}(r)$ in Eqs. (\ref{a3}) and (\ref{a5}), a nice cancellation of terms occurs, leaving us with
\begin{equation}
\label{a9}
\rho(r)=\frac{m_B^2}{4\pi^2\hbar}\int d\epsilon dl\, (2l+1) |{\tilde a}_{nl}|^2\frac{1}{r^2p(r)}.
\end{equation}
For a given $\epsilon$ and $r$, $l$ ranges from $0$ to $l_{\rm max}$ such that $\epsilon-l(l+1)\hbar^2/(2m_B^2r^2)-V=0$. If we assume that $ |{\tilde a}_{nl}|^2$ depends only on $\epsilon=E_{nl}/m_B$ (in agreement with the corresponding assumption that $f=f(\epsilon)$ in the classical description) the integral over $l$ can be easily performed with the change of variables $x=l(l+1)$ yielding
\begin{equation}
\label{a10}
\rho(r)=\frac{m_B^3}{2\pi^2\hbar^3}\int d\epsilon  |{\tilde a}_{nl}|^2 \sqrt{2(\epsilon-V)}.
\end{equation}
Comparing Eqs. (\ref{a2}) and (\ref{a10}) the following relation is obtained \cite{YavetzLiHui2022}:
\begin{equation}
\label{a11}
f(\epsilon)=\frac{m_B^3}{(2\pi \hbar)^3}|{\tilde a}_{nl}|^2.
\end{equation}

This equality is approximate in the sense that it is valid in the WKB limit. It is expected to hold only for eigenmodes with a high enough energy $\epsilon$, i.e., for eigenmodes that describe the envelope of the DM halo. The soliton has to be treated independently as being the ground state of the SP equations. The interface between the soliton and the halo (with intermediate energies) may not be accurately described by the WKB approximation.

In conclusion, for a spherically symmetric halo with a particle distribution function $f(\epsilon)$, the density profile is given by Eq. (\ref{a2}) and the wave is given by Eq. (\ref{psi0}) with $|{\tilde a}_{nl}|^2$ given by Eq. (\ref{a11}) in the WKB limit, i.e., for large energies. Using this kind of construction \cite{PhysRevD.97.103523} have shown that the time-average envelope of FDM halos obtained in numerical simulations is well-fitted by the fermionic King model \cite{c1998} giving further support to the claim made in \cite{clm2015} that the (fermionic) King model may be a good model of the envelope of DM halos.

\subsection{WKB for functionals}

Using the WKB approximation for $R_{nl}(r)$ [see Eqs. (\ref{a3}) and (\ref{a5})], and proceeding as above, we find that the energy functional defined by Eq. (\ref{ber1}) reduces to 
\begin{eqnarray}
\label{a13}
E=\frac{2m_B^3}{\pi\hbar^3}\int d\epsilon  |{\tilde a}_{nl}|^2 \sqrt{2(\epsilon-V)}\epsilon r^2\, dr.
\end{eqnarray}
Using the identity from Eq. (\ref{a11}) it can be rewritten as
\begin{eqnarray}
\label{a14}
E=16\pi^2\int d\epsilon  f(\epsilon)\epsilon \sqrt{2(\epsilon-V)} r^2\, dr
\end{eqnarray}
or as
\begin{eqnarray}
\label{a15}
E=\int  f(\epsilon)\epsilon \, d{\bf r}d{\bf v}.
\end{eqnarray}
Recalling that $\epsilon=v^2/2+V$ we obtain
\begin{eqnarray}
\label{a16}
E=\int  f\frac{v^2}{2} \, d{\bf r}d{\bf v}+2W.
\end{eqnarray}
Finally, recalling the identity $E=K+2W$ established in Sec. \ref{sec:ba} and using Eq. (\ref{a16}),  we find that the quantum kinetic energy coincides, in the high energy limit, with the classical kinetic energy
\begin{eqnarray}
\label{a17}
K=\int  f\frac{v^2}{2} \, d{\bf r}d{\bf v}.
\end{eqnarray}
This agreement is expected, but not trivial, since $K$ in Eq. (\ref{eq:K0}) is expressed in terms of the wave function $\psi({\bf r},t)$ -- a function of position only -- while $K$ in Eq. (\ref{a17}) is expressed in terms of the DF $f({\bf r},{\bf v},t)$ -- a function of position and velocity (reducing to a function of the energy $\epsilon$ for spherically symmetric systems). 

Finally, we emphasize that the total energy (the one which is conserved) is $E_{\rm tot}=K+W$. It differs from the energy $E$ (related to the eigenenergies) which is given by $E=K+2W$. The factor $2$ arises because the system is self-gravitating (instead of being subjected to an external potential).

\bibliography{BECDM}

\end{document}